\documentclass[fleqn,usenatbib]{mnras}
\usepackage[T1]{fontenc}
\DeclareRobustCommand{\VAN}[3]{#2}
\let\VANthebibliography\thebibliography
\def\thebibliography{\DeclareRobustCommand{\VAN}[3]{##3}\VANthebibliography}
\usepackage{graphicx}	
\usepackage{amssymb}	
\usepackage{lscape}
\usepackage{ulem}
\usepackage{tablefootnote}
\usepackage{lipsum}
\usepackage{schemata}

\title[Globular clusters' rotation-mass relation]{First observational evidence of a relation between globular clusters' internal rotation and stellar masses}

\author[Scalco et al.]{M.\,Scalco$^{1,2,3}$\thanks{E-mail: michele.scalco@unife.it},
A.\,Livernois$^3$, 
E.\,Vesperini$^3$,
M.\,Libralato$^4$,
A.\,Bellini$^5$ and
L.\,R.\,Bedin$^2$
\\
$^{1}$Dipartimento di Fisica e Scienze della Terra, Università di Ferrara, Via Giuseppe Saragat 1, Ferrara I-44122, Italy\\
$^{2}$Istituto Nazionale di Astrofisica, Osservatorio Astronomico di Padova, Vicolo dell’Osservatorio 5, Padova I-35122, Italy\\
$^{3}$Department of Astronomy, Indiana University, Bloomington, Swain West, 727 E. 3rd Street, IN 47405, USA\\
$^{4}$AURA for the European Space Agency (ESA), Space Telescope Science Institute, 3700 San Martin Drive, Baltimore, MD 21218, USA\\ 
$^{5}$Space Telescope Science Institute, 3700 San Martin Drive, Baltimore, MD 21218, USA\\
}

\date{Accepted 2023 March 22. Received 2023 March 21; in original form 2023 January 25}

\pubyear{2023}

\begin{document}
\label{firstpage}
\pagerange{\pageref{firstpage}--\pageref{lastpage}}
\maketitle

\begin{abstract}
Several observational studies have shown that many Galactic globular clusters (GCs) are characterised by internal rotation. Theoretical studies of the dynamical evolution of rotating clusters have predicted that, during their long- term evolution, these stellar systems should develop a dependence of the rotational velocity around the cluster’s centre on the mass of stars, with the internal rotation increasing for more massive stars. In this paper we present the first observational evidence of the predicted rotation-mass trend. In our investigation, we exploited the \textit{Gaia} Data Release 3 catalogue of three GCs: NGC 104 (47 Tuc), NGC 5139 ($\omega$ Cen) and NGC 5904 (M 5). We found clear evidence of a cluster rotation-mass relation in 47 Tuc and M 5, while in $\omega$ Cen, the dynamically youngest system among the three clusters studied here, no such trend was detected.
\end{abstract}


\section{Introduction} \label{sect:intro}
High-precision proper-motion (PM) studies enabled by space-based facilities such as \textit{Hubble Space Telescope} (\textit{HST}) and \textit{Gaia}, and line-of-sight velocities produced by large ground-based spectroscopic surveys are providing key insights into the internal kinematic properties of GCs and have opened a new important window in the study of these systems. The results emerging from these observational investigations are revealing a dynamical picture that differs significantly from the traditional view of GCs as isotropic, non-rotating systems.\looseness=-4

Numerous studies have now found that internal rotation is a common feature in GCs (e.g., \citealt{2017ApJ...844..167B}, \citealt{2018MNRAS.481.2125B}, \citealt{2018ApJ...860...50F}, \citealt{2018ApJ...861...16L}, \citealt{2018MNRAS.480.1689K}, \citealt{2021MNRAS.505.5978V}) and several clusters for which observations allowed to investigate the kinematic properties in a broad range of radial distances from the clusters' centres have often revealed an anisotropic velocity distribution in the clusters' outer regions (e.g., \citealt{2014ApJ...797..115B}, \citealt{2015ApJ...803...29W}, \citealt{2019MNRAS.487.3693J}, \citealt{2022ApJ...934..150L}).\looseness=-4

The discovery of multiple stellar populations in GCs (see e.g. \citealt{2019A&ARv..27....8G} for a recent review) has provided evidence of further complexities in various aspects of the study of these systems and a few early studies have shown differences in the rotation and anisotropy of different stellar populations (e.g., \citealt{2013ApJ...771L..15R}, \citealt{2017MNRAS.465.3515C}, \citealt{2018ApJ...853...86B}, \citealt{2020ApJ...889...18C,2020ApJ...898..147C}, \citealt{2020MNRAS.492.2177K}, \citealt{2021MNRAS.506..813D}, \citealt{2023ApJ...944...58L}).\looseness=-4

The present-day dynamical properties of GCs provide important information both on the initial conditions imprinted at the time of the cluster's formation and on the effects of long-term evolutionary processes such as internal two-body relaxation and the interactions with the external tidal field of the Galaxy. Effects of dynamical evolution on the internal kinematics include, for example: the gradual development of a dependence of the velocity dispersion on the stellar mass as clusters evolve towards energy equipartition (\citealt{2013MNRAS.435.3272T}), a radial anisotropy in the cluster's velocity distribution --- which may be imprinted in various evolutionary phases and whose variation with the clustercentric distance is subsequently affected by internal relaxation, stellar escape, and the external tidal field (see e.g. \citealt{2005A&A...433...57T}, \citealt{2014MNRAS.443L..79V}, \citealt{2022MNRAS.512.1584T}).\looseness=-4

As mentioned above, many GCs are characterised by internal rotation. Numerical simulations (see e.g. \citealt{1999MNRAS.302...81E}, \citealt{2007MNRAS.377..465E}, \citealt{2017MNRAS.469..683T}, \citealt{2022MNRAS.516.3266K}) have shown that internal relaxation and stellar escape lead to a redistribution and loss of angular momentum so that the strength of the present-day rotation is expected to be smaller than the primordial one emerging at the end of the formation phase.\looseness=-4

As shown in a number of theoretical studies, the presence of internal rotation has several implications for the long-term dynamical evolution of clusters and the rotational properties of clusters may shed light on a number of fundamental aspects of the dynamics of GCs. Internal rotation may affect the rate of stellar escape and the cluster's evolution towards core collapse (see e.g. \citealt{2022MNRAS.516.3266K} and references therein), the coupling between internal rotation and the cluster's orbital motion can lead to a variation of the orientation of the internal rotation axis with the clustercentric distance \citep{2018MNRAS.475L..86T,2022MNRAS.512.1584T}. Numerical simulations have also shown that internal rotation has a significant impact on the fundamental effects of two-body relaxation; specifically, a number of studies have found that rotating star clusters are characterised by anisotropic mass segregation (see e.g. \citealt{2019ApJ...887..123S}, \citealt{2021MNRAS.506.5781L,2022MNRAS.512.2584L}, \citealt{2021MNRAS.506.4488T}) and a difference between the degree of energy equipartition in the tangential and radial components of the velocity dispersion (see e.g. \citealt{2022MNRAS.512.2584L}).\looseness=-4

Numerical simulations have also predicted that, during a cluster's evolution, the rotational velocity develops a dependence on the stellar mass with more massive stars tending to rotate more rapidly than low-mass stars (see \citealt{2004MNRAS.351..220K}, \citealt{2013MNRAS.430.2960H}, \citealt{2021MNRAS.506.5781L,2022MNRAS.512.2584L}). In this Letter, we focus our attention on this prediction and present an analysis, based on \textit{Gaia} Data Release 3 (\textit{Gaia} DR3, \citealt{2016A&A...595A...2G,2022arXiv220800211G}) data of three GCs (47 Tuc, $\omega$ Cen and M 5), providing the first observational evidence of the predicted trend between rotational velocity and stellar mass.\looseness=-4

\begin{table*}
\caption{Proprieties of the three studied clusters. For each cluster, the table provides: name of the cluster, position and half-light radius, $r_{\rm h}$, from \citet{1996AJ....112.1487H,2010arXiv1012.3224H}, mean PM ($\langle \mu_{\alpha}cos\delta \rangle$,$\langle \mu_{\delta} \rangle$) and central velocity dispersion, $\sigma_0$, from \citet{2021MNRAS.505.5978V}, radius used in this work for the data extraction from the \textit{Gaia} DR3 catalogue, $r_{\rm ext}$, number of stars passing the astrometric and photometric quality selections, $N_{\rm s}$, and number of selected member stars, $N_{\rm m}$.}
\centering
\begin{tabular}{l c c c c c c c c c}
    \hline
    \hline
{\footnotesize ID} & {\footnotesize R.A. (J2000)} & {\footnotesize Decl. (J2000)} & {\footnotesize $r_{\rm h}$} & {\footnotesize $\langle \mu_{\alpha}cos\delta \rangle$} & {\footnotesize $\langle \mu_{\delta} \rangle$} & {\footnotesize $\sigma_0$} & {\footnotesize $r_{\rm ext}$} & {\footnotesize $N_{\rm s}$} & {\footnotesize $N_{\rm m}$}\\
 & {\footnotesize [h m s]} & {\footnotesize [$^\circ$ $^\prime$ $^{\prime \prime}$]} & {\footnotesize [arcmin]} & {\footnotesize [mas/yr]} & {\footnotesize [mas/yr]} & {\footnotesize [mas/yr]} & {\footnotesize [arcmin]} & &\\
\hline
{\footnotesize NGC 104} & {\footnotesize 00 24 05.67} & {\footnotesize $-$72 04 52.6} & {\footnotesize 3.17} & {\footnotesize 5.252 $\pm$ 0.021} & {\footnotesize $-$2.551 $\pm$ 0.021} & {\footnotesize 0.5463} & {\footnotesize 40} & {\footnotesize 61257} & {\footnotesize 36608}\\
{\footnotesize NGC 5139} & {\footnotesize 13 26 47.24} & {\footnotesize $-$47 28 46.5} & {\footnotesize 5.00} & {\footnotesize $-$3.250 $\pm$ 0.022} & {\footnotesize $-$6.746 $\pm$ 0.022} & {\footnotesize 0.5610} & {\footnotesize 40} & {\footnotesize 92603} & {\footnotesize 45665}\\
{\footnotesize NGC 5904} & {\footnotesize 15 18 33.22} & {\footnotesize $+$02 04 51.7} & {\footnotesize 1.77} & {\footnotesize 4.086 $\pm$ 0.023} & {\footnotesize $-$9.870 $\pm$ 0.023} & {\footnotesize 0.2343} & {\footnotesize 25} & {\footnotesize 10613} & {\footnotesize 6344}\\
     \hline
  \end{tabular}
\label{Table1}
\end{table*}

\section{Gaia DR3 data: selections and cluster membership}\label{sec:gaia}

Our analysis is specifically focused on three GCs (47 Tuc, $\omega$ Cen, and M 5). These three clusters have been selected since they are among those with the strongest internal rotation and their distance from the Sun is such that the range of stellar magnitudes available from the \textit{Gaia} data corresponds to a stellar mass range sufficient ($\sim$0.6--0.85 M$_{\odot}$) to explore the dependence of rotation on the stellar mass.\looseness=-4

For each of the clusters analysed in our study, we retrieved astrometry, photometry, parallaxes ($\pi$), and PMs of all sources from the \textit{Gaia} DR3 archive that are located within a given radius from the nominal cluster centre (the exact value of this radius and the central coordinate for each cluster are reported in Table \ref{Table1}). We defined a sample of stars with reliable photomety and astrometry by following the recommendations of \citet{2021A&A...649A...5F}, \citet{2021A&A...649A...2L} and \citet{2021A&A...649A...3R}, but with tighter selections on some parameters (see Appendix in the online material for details on the specific selection applied). The number of stars passing the photometric and astrometric quality selections is reported in Table \ref{Table1} for each cluster.\looseness=-4

\subsection{Selection of cluster members}
\begin{figure}
\centering
 \includegraphics[width=\columnwidth]{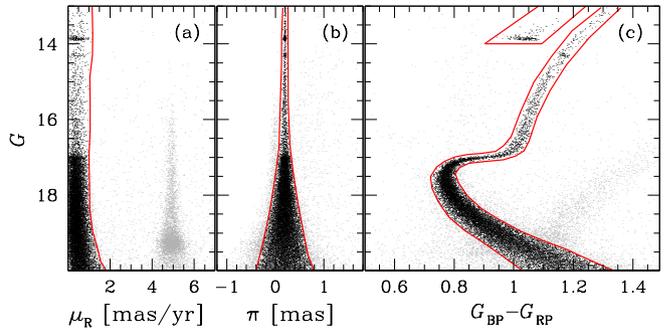}
 \caption{This figure illustrates the procedure used to select probable members of 47 Tuc. (a) $G$ versus $\mu_{\rm R}$; (b) $G$ versus $\pi$; (c) $G$ versus $G_{\rm BP}-G_{\rm RP}$ CMD. The red lines in all panels are used to separate cluster members from field stars. Selected cluster members are in black, while field stars are in grey. Only the 60\% of the sample is shown for clarity.\looseness=-4}
 \label{Figure2} 
\end{figure}

To derive the clusters rotation from the \textit{Gaia} DR3 catalogue, we first need to select the most probable cluster members for each cluster. To do this, we used a procedure based on stellar PMs, $\pi$ and positions in the colour-magnitude diagram (CMD). The procedure is illustrated in Fig. \ref{Figure2} for 47 Tuc and summarised in the following:\looseness=-4
\begin{enumerate}
    \item[1.] We initially applied the orthographic projection of the celestial coordinates and converted PMs using Equation (2) from \citet{2018A&A...616A..12G}. We plotted $G$ as a function of the PM relative to the mean PM of the cluster \citep[from][see also Table \ref{Table1}]{2021MNRAS.505.5978V}, $\mu_{\rm R}$, (panel a of Fig. \ref{Figure2}) for each source in the sample. We drew by hand a line that separates cluster stars ($\mu_{\rm R}$ close to 0) from field stars, and defined a sample of probable cluster members.\looseness=-4
    \item[2.] We plotted $G$ as a function of $\pi$ for the probable cluster members, and drawn by hand two lines that enclose all the stars with likely cluster $\pi$ (panel b of Fig. \ref{Figure2}). All the stars that lie outside these two lines were excluded from the sample of probable cluster members.\looseness=-4
    \item[3.] We plotted the $G$ versus $G_{\rm BP}-G_{\rm RP}$ CMD for the probable cluster members (panel c of Fig. \ref{Figure2}). We identified in the cleaned CMD the sequence of stars associated to the cluster and defined by hand lines that follow the red and blue edges of the sequence profiles. We then removed from the sample of probable cluster members all the stars that lie outside of these boundaries.\looseness=-4
\end{enumerate}
We iterate this procedure a few times to improve our membership selection. The number of cluster members for each GC and other physical parameters are reported in Table \ref{Table1}.\looseness=-4

\section{Globular clusters rotation} \label{sec:results}
In this section, we briefly review the theoretical predictions about the dependence of the cluster internal rotational velocity on the stellar mass, and then present the observational results obtained from the analysis of the \textit{Gaia} DR3 catalogue.\looseness=-4

\subsection{Theoretical predictions}\label{theoretical}
A few theoretical studies on the dynamics of rotating star clusters  \citep{2004MNRAS.351..220K,2013MNRAS.430.2960H,2021MNRAS.506.5781L,2022MNRAS.512.2584L} have found that one of the fundamental effects of dynamical evolution of rotating star clusters is the development of a dependence of the internal rotation on the mass of the stars, where the rotational velocity increases with the stellar mass. We refer to those papers for a complete description of the evolutionary path of the various models explored. Here we focus our attention on the N-body models discussed in \citet{2022MNRAS.512.2584L}.\looseness=-4

Although these models are not aimed at a direct comparison with observational data of any specific GC, in order to establish a closer connection with the analysis presented in this paper, we quantify the development and the strength of the dependence of the rotational velocity on the stellar mass found in the Livernois et al's models using the same parameters adopted in our observational analysis.\looseness=-4

\begin{figure}
\centering
\includegraphics[width=0.49\columnwidth]{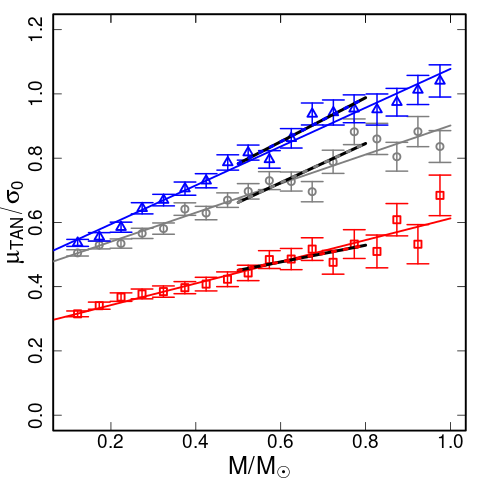}
\includegraphics[width=0.49\columnwidth]{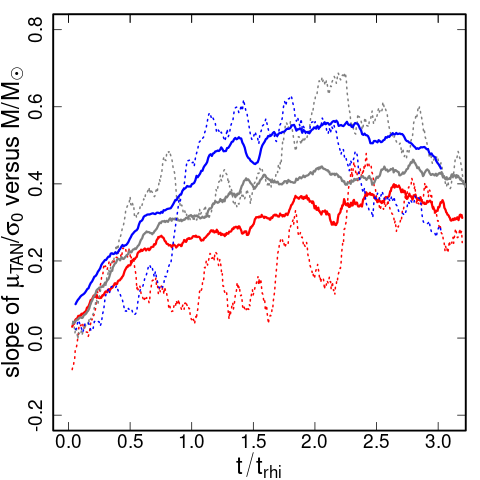}
 \caption{\textit{Left panel}: $\mu_{\rm TAN}/\sigma_0$ versus mass measured at $t/t_{\rm rhi}$ = 2 for for the low (red), moderate (grey), and high (blue) rotation models studied in \citet{2022MNRAS.512.2584L}. Thin (thick) lines show the best linear fit using the 0.1--1 M$_{\odot}$ (0.5--0.8 M$_{\odot}$) mass range. \textit{Right panel}: slope of the $\mu_{\rm TAN}/\sigma_0$-mass relationship as a function of time (normalised to $t_{\rm rhi}$) for all the three models. The continuous lines represent the running average of the slope calculated using the full mass range available in the simulation (0.1--1 M$_{\odot}$), while the dashed lines represent the running average of the slope calculated using a narrower mass range closer to that available in the observational analysis (0.5--0.8 M$_{\odot}$).\looseness=-4}
 \label{Figure3} 
\end{figure}

\citet{2022MNRAS.512.2584L} studied the dynamical evolution of three multi-mass, rotating star cluster models characterised by low-, moderate-, and high-rotation; in the left panel of Fig. \ref{Figure3}, we plot $\mu_{\rm TAN}/\sigma_0$ (where $\mu_{\rm TAN}$ is the tangential component of the velocity and $\sigma_0$ is the central velocity dispersion of all stars, evaluated on the plane perpendicular to the rotation axis) versus mass measured at $t/t_{\rm rhi}$ = 2 (where $t_{\rm rhi}$ is the initial half-mass relaxation time) for those three models. It is interesting to point out that the models predict a linear relationship between $\mu_{\rm TAN}/\sigma_0$ and stellar mass and its strength is thus properly quantified by the slope shown in this panel. In the right panel, we plot the time evolution of the slope of the $\mu_{\rm TAN}/\sigma_0$-mass relationship as a function of time, for all three models.\looseness=-4

As shown in this figure, all models develop a mass-dependent rotation curve that is characterised by a rotation increasing with the stellar mass. The slope of the $\mu_{\rm TAN}/\sigma_0$-mass relationship depends on the strength of the initial rotation and is slightly steeper for models with more rapid initial rotation. In this figure we show both the slope calculated using the full mass range available in the simulation (0.1--1 M$_{\odot}$) and the same slope calculated using a narrower mass range closer to that available in the observational analysis (0.5--0.8 M$_{\odot}$; see section \ref{observational}). As expected, the calculation using a broader mass range results in a more robust and less noisy estimate of the slope of the $\mu_{\rm TAN}/\sigma_0$-mass relationship but both estimates follow a similar time evolution.\looseness=-4

\subsection{Observational results}\label{observational}
In light of the results presented in \citet{2022MNRAS.512.2584L} and shown in Fig. \ref{Figure3} we made use of our \textit{Gaia} DR3 catalogue to investigate the presence of a rotation-mass relation in our three GCs, 47 Tuc, $\omega$ Cen and M 5.\looseness=-4

\begin{centering} 
\begin{figure*}
\centering
 \includegraphics[width=5cm]{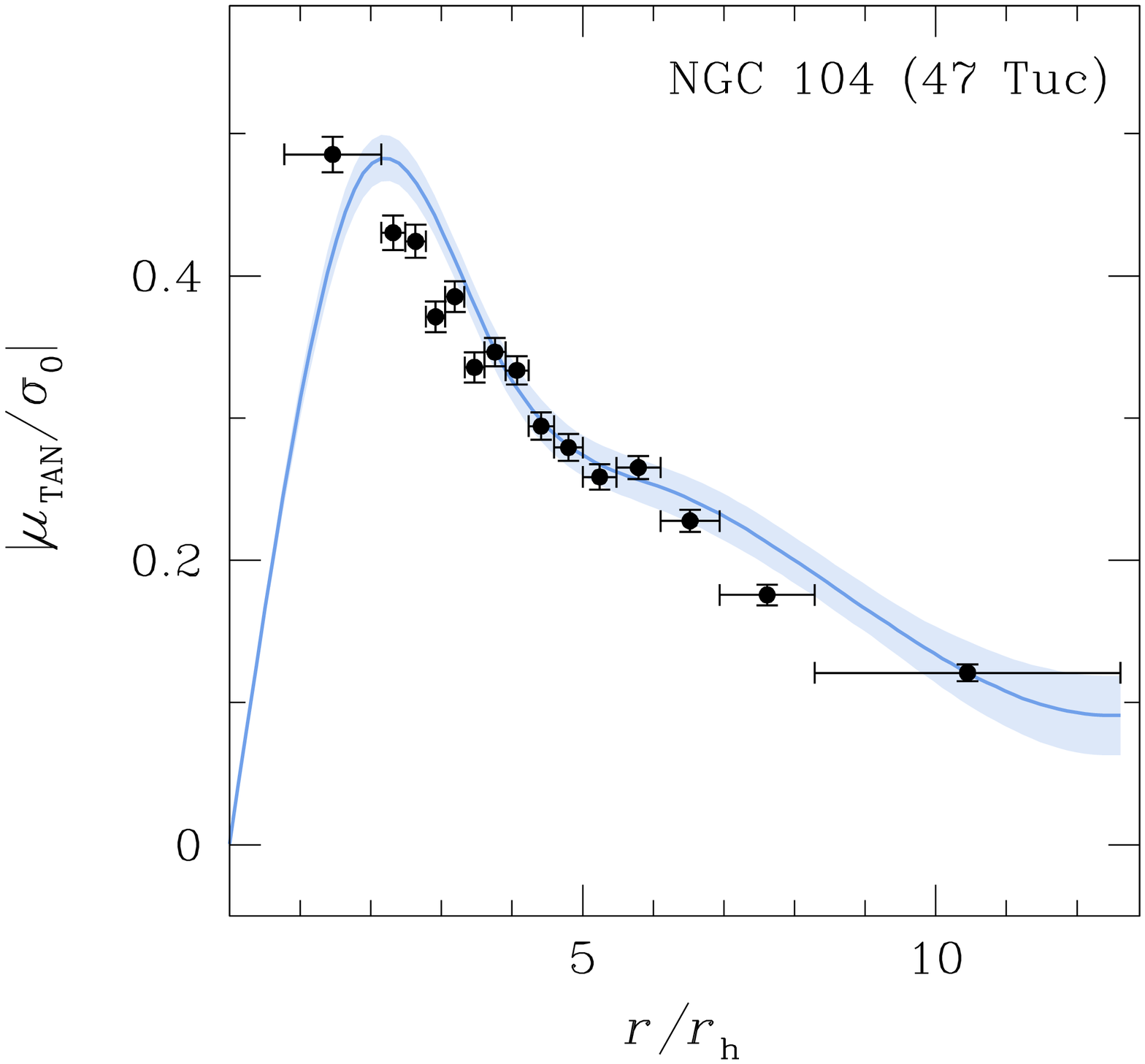}
 \includegraphics[width=5cm]{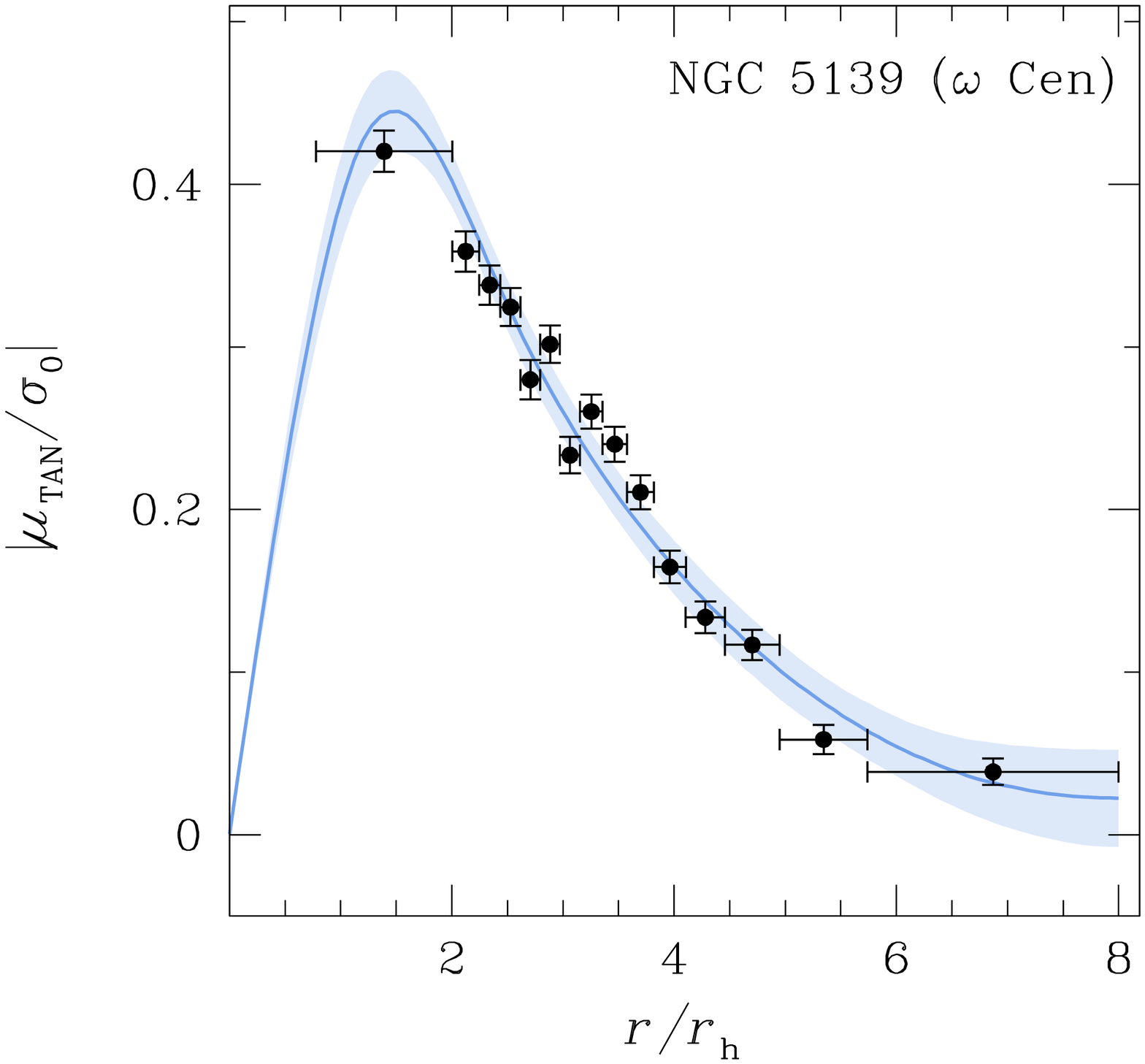}
 \includegraphics[width=5cm]{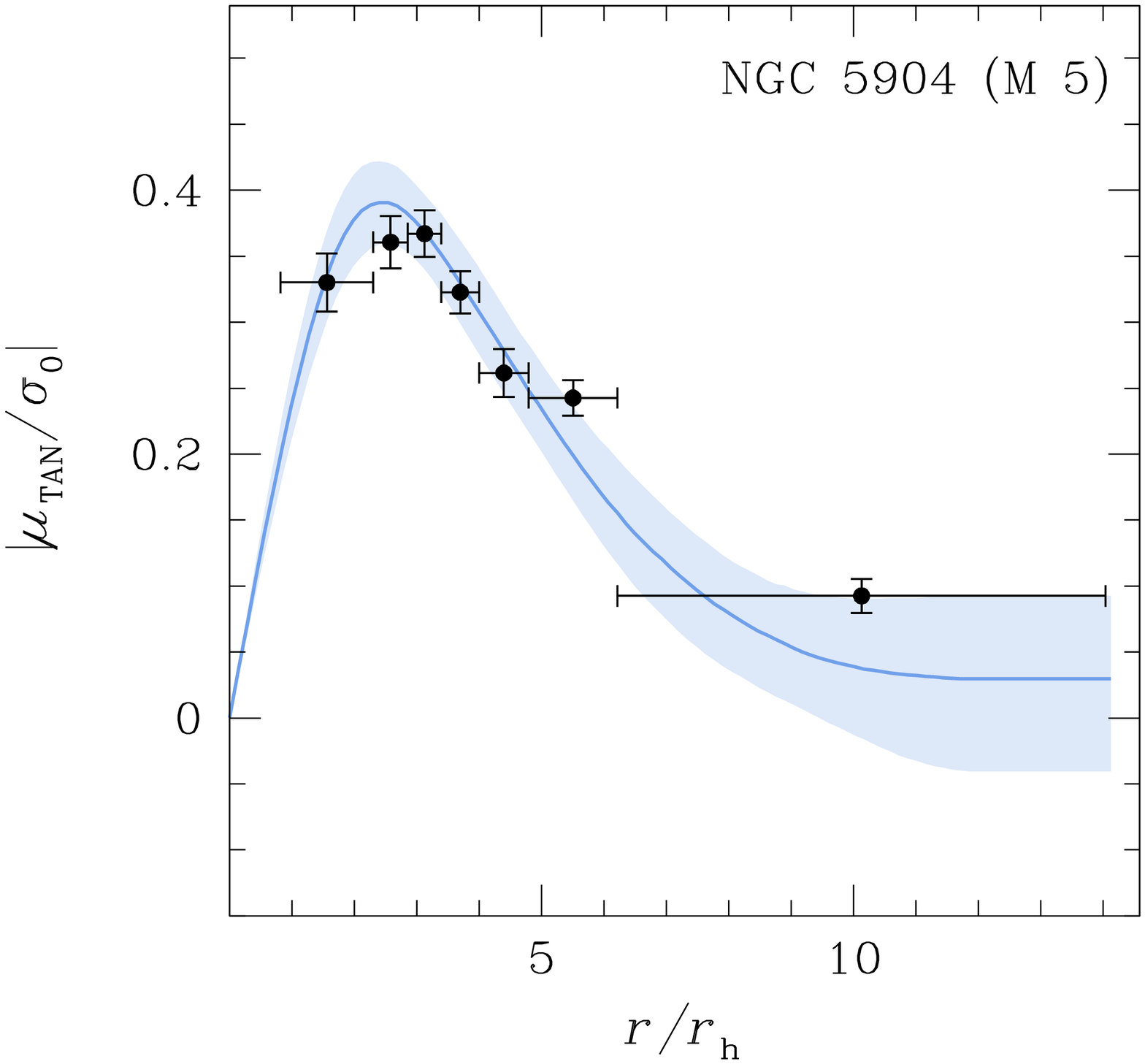}
 \includegraphics[width=5cm]{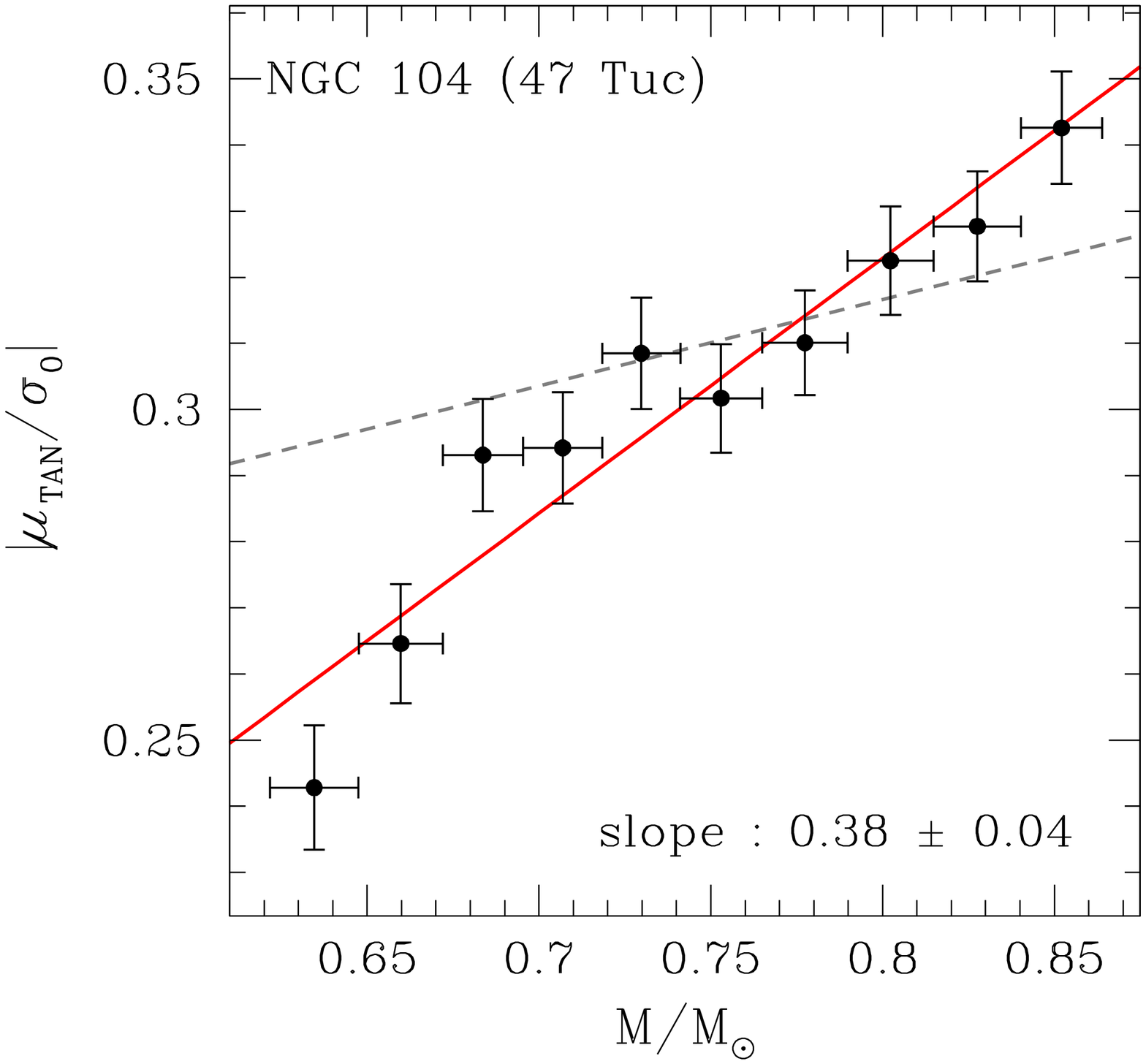}
 \includegraphics[width=5cm]{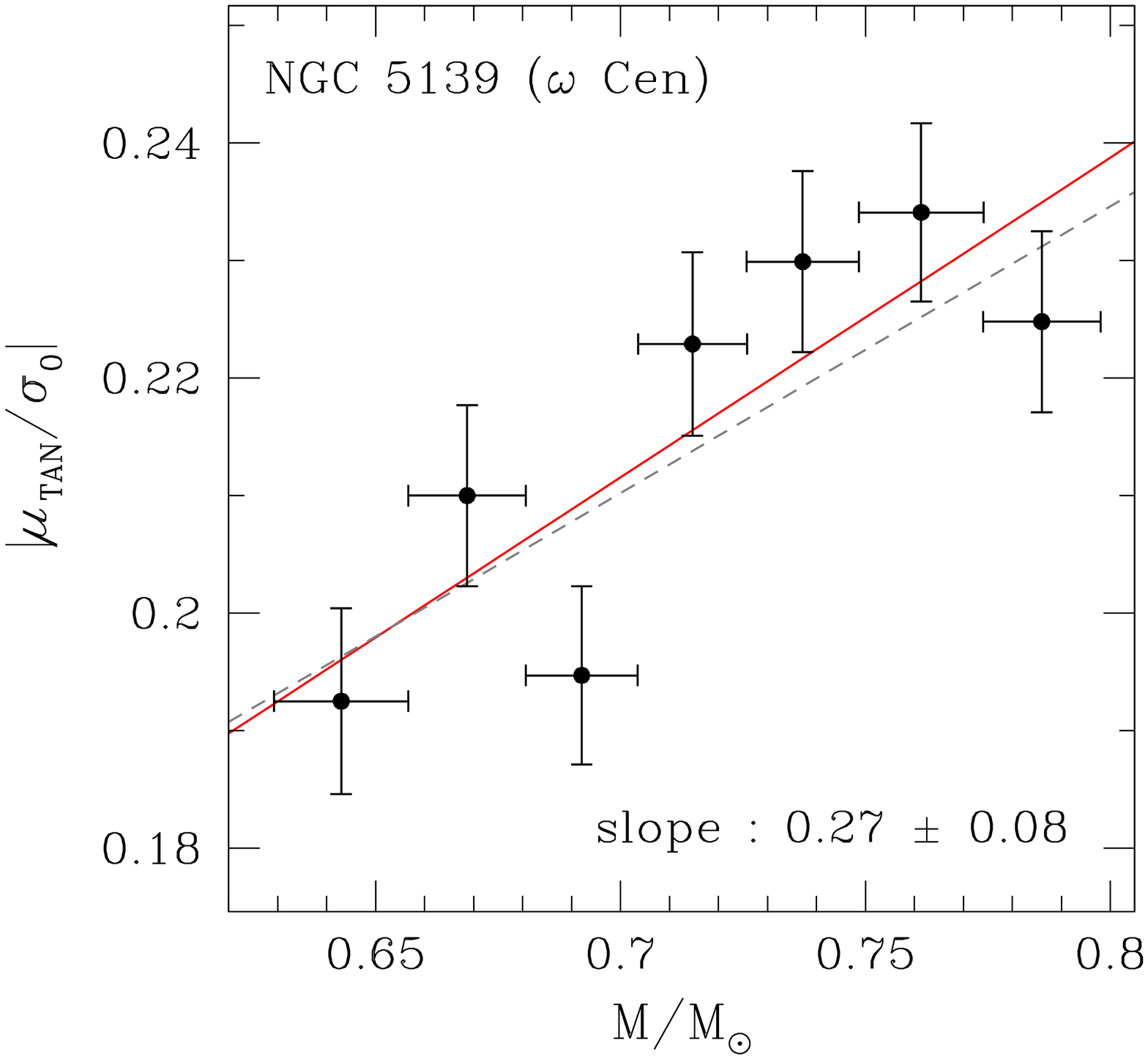}
 \includegraphics[width=5cm]{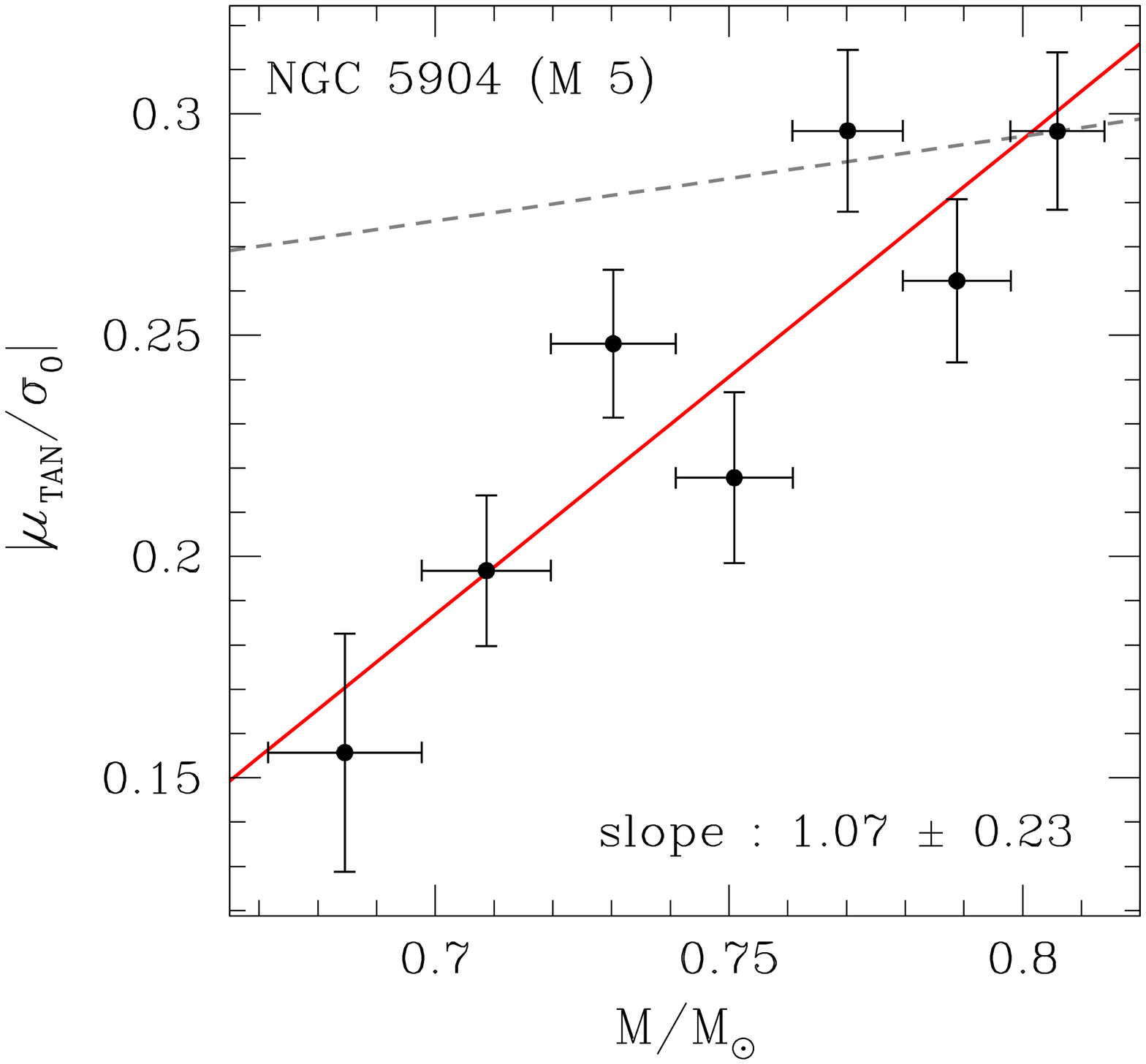}
 \caption{\textit{Top panels}: absolute value of the rotation profile in the plane of the sky obtained with the sample of cluster members selected in this work from the \textit{Gaia} DR3 catalogue (black points), for the three clusters. The rotation curves are normalised by the central velocity dispersion value, $\sigma_{0}$, provided by \citet{2021MNRAS.505.5978V}. $r_{\rm h}$ is the half-light radius of the cluster from \citet{1996AJ....112.1487H,2010arXiv1012.3224H}. The values of $\sigma_0$ and $r_{\rm h}$ are reported in Table \ref{Table1}, for each cluster. The solid blue lines and shaded areas represent the rotation profiles presented in \citet{2021MNRAS.505.5978V} and their 68\% confidence intervals, respectively. \textit{Bottom panels}: $|\mu_{\rm TAN}/\sigma_{0}|$ as a function of the stellar mass (black points). The red lines represent the least square fits of the bins. The slopes of the red lines are also reported. The dashed lines show the slopes obtained with the test described in section \ref{observational} to determine the mass-rotational velocity trend due only to mass segregation and radial variation of the rotation curve  ($0.13\pm0.01$ for 47 Tuc, $0.24\pm0.03$ for $\omega$ Cen and $0.19\pm0.02$ for M 5)\looseness=-4}
 \label{Figure4} 
\end{figure*} 
\end{centering}

We initially decomposed the PMs into radial and tangential components ($\mu_{\rm RAD}$, $\mu_{\rm TAN}$) and propagated the velocity uncertainties. For each cluster, we divided the sample of member stars in equally populated radial bins. Then, for each bin, we evaluated the  mean value of $\mu_{\rm TAN}$, using a maximum likelihood approach \citep{2022ApJ...934..150L}. The top panels of Fig. \ref{Figure4}, show the obtained rotation profiles, in absolute value, for each cluster (black points). The profiles are normalised by the central velocity dispersion value, $\sigma_{0}$, provided by \citet{2021MNRAS.505.5978V} and reported in Table \ref{Table1}. The rotation curves are compared with those published in \citet{2021MNRAS.505.5978V} (blue solid line), showing a quite good agreement.\looseness=-4

We converted magnitudes into stellar masses via isochrone fitting, for which we used isochrones from the Dartmouth Stellar Evolution Database \citep{2008ApJS..178...89D}. We adopted cluster metallicities [Fe/H] from \citet{1996AJ....112.1487H,2010arXiv1012.3224H} and assumed primordial helium abundance ($Y$ = 0.246). We assumed [$\alpha$/Fe] = $+$0.2 for 47 Tuc and M 5 \citep{2010ApJ...708..698D}, and [$\alpha$/Fe] = $+$0.4 for $\omega$ Cen \citep{2020MNRAS.497.3846M}, and adopted age, distance modulus and reddening that best fit the data. We then associated to each star the corresponding mass in the interpolating isochrone\footnote{We neglected the presence of multiple stellar-populations since their effect is likely to be small, as shown in \citet{2019ApJ...873..109L}.}. The reddening has been converted into absorption using the extinction coefficient provided in \citet{2018MNRAS.479L.102C}. Following \citet{2022ApJ...936..154W}, we identified in the CMD all the stars belonging to the AGB and HB and reassign their masses to be equal to the maximum in the sample. We limited our analysis to stars located at a distance $r>2r_{\rm h}$ so to remove the inner radial regions affected by a significant incompleteness in the \textit{Gaia} data after the selections (see Appendix in the online material).\looseness=-4
Finally, for each cluster we divided the sample of member stars equally populated mass bins. Then, for each bin, we evaluated the mean value of $\mu_{\rm TAN}$, again with the maximum likelihood method. The absolute values of the mean rotational velocities, normalised by $\sigma_0$, are shown in the bottom panels of Fig. \ref{Figure4} as a function of the stellar mass, along with the least square linear fit (red lines). As shown by this Figure, there is a clear trend of rotational velocity increasing with the stellar mass.\looseness=-4
It is important to point out that such a trend could also result from the combined effect of mass segregation (massive stars preferentially populating the inner regions) and the increase of  the rotational velocity at smaller distances from the cluster's centre. In order to estimate the extent of this effect we have assigned to each star a value of rotational velocity depending only on the star's radial position (and calculated from the rotational velocity profile calculated in our analysis and shown in top panels of Fig. \ref{Figure4}). In this test, all stars at a given distance from the cluster's centre have the same rotational velocity regardless of their stellar mass: by construction, we have completely removed the specific mass-dependence of $\mu_{\rm TAN}$ predicted by simulations and on which our study is focused. With these data, we have then calculated the $\mu_{\rm TAN}/\sigma_0$-mass slope which, in this case, is entirely due to the combined effect of mass segregation and radial variation of $\mu_{\rm TAN}$. The slope of the $\mu_{\rm TAN}$-mass trend obtained in this test is $0.13\pm0.01$ for 47 Tuc and $0.19\pm0.02$ for M 5; in both cases the slope is significantly weaker than that found in the full analysis and confirms that trend found in our study for these two clusters is due to the effect predicted by theoretical studies. For $\omega$ Cen, on the other hand, the slope found with this test is equal to $0.24\pm0.03$ and is consistent with that obtained in the full analysis showing that for this cluster no significant $\mu_{\rm TAN}$-mass trend (other than that simply due to mass segregation and radial variation of $\mu_{\rm TAN}$) is detected.\looseness=-4

It is interesting to point out that that the lack of a significant $\mu_{\rm TAN}$-mass trend in $\omega$ Cen might be due to the fact that this cluster is the dynamically youngest system among the three clusters studied (the half-mass relaxation time is $\log$($t_{\rm rh}$/yr) $=$ 10.39, 9.58, and 9.45 for $\omega$ Cen, 47 Tuc, and M 5, respectively; \citealt{2018MNRAS.478.1520B}) and such a trend might not have developed yet (or still be very weak).\looseness=-4

As for the slopes found in 47 Tuc and M 5, we notice that the slope for 47 Tuc is consistent with those found in our simulations while the slope for M 5 is a bit larger than those in our models. As pointed out in section \ref{theoretical}, the goal of these simulations is not a close comparison with observational data but rather that of illustrating the fundamental effects of dynamical evolution on the development of a $\mu_{\rm TAN}$-mass trend. Simulations spanning a broader range of initial structural and kinematic initial conditions would be necessary for a full exploration of the possible range of values of the $\mu_{\rm TAN}$-mass slopes.\looseness=-4

\section{Conclusion} \label{sec:conclusion}
At odds with the traditional dynamical picture according to which GCs would be isotropic and non-rotating stellar systems, several recent observational studies have shown that many GCs are instead characterised by the presence of internal rotation. A few theoretical studies have focused their attention on the dynamics of rotating clusters and explored the effects of various dynamical processes on the evolution of a cluster's internal rotation. One of the predictions of these theoretical studies is that, during their long-term dynamical evolution, GCs develop a dependence of their internal rotational velocity on the stellar mass where more massive stars tend to rotate around the cluster's centre more rapidly than low-mass stars.\looseness=-4

In this letter, after reviewing the theoretical predictions using the simulations of rotating clusters of \citet{2022MNRAS.512.2584L}, we have exploited data from the \textit{Gaia} DR3 catalogue to carry out an analysis of the rotational properties of three GCs (47 Tuc, $\omega$ Cen, and M 5) aimed at validating the predicted relation between internal rotation about the cluster's centre and stellar mass.\looseness=-4

Our study reveals that 47 Tuc and M 5 are characterized by a trend between rotation and stellar mass, where the rotational velocity increases with the stellar mass and provide the first observational evidence of the predicted rotation-mass relation. The trend between rotational velocity and mass found in $\omega$ Cen, on the other hand, is consistent with being simply due to the combined effect of mass segregation and radial variation of the rotational velocity; $\omega$ Cen is a dynamically young cluster, and the lack of a significant $\mu_{\rm TAN}$-mass (or the presence of a weaker one undetected in our analysis) trend is generally consistent with the expected development of this relation during a cluster's long term-evolution.\looseness=-4

Due to the faint magnitude limit of \textit{Gaia}, we focused our analysis only on nearby and rapidly rotating GCs, in order to have a sufficient mass interval and rotation signal to properly characterise the $\mu_{\rm TAN}/\sigma_0$-mass slope. Future observational investigations, possibly based on \textit{HST}, \textit{JWST})and \textit{Roman Space Telescope} PM measurements, extending the mass range down to lower masses than those currently available, and out to the tidal radius, would significantly strengthen the investigation of the $\mu_{\rm TAN}/\sigma_0$-mass relationship and provide the opportunity to carry out a comprehensive investigation of this trend, explore the possible dependence of its strength on other clusters' dynamical properties and evolutionary history.\looseness=-4

\section*{Acknowledgements}
This work has made use of data from the European Space Agency (ESA) mission
{\it Gaia} (\url{https://www.cosmos.esa.int/gaia}), processed by the {\it Gaia}
Data Processing and Analysis Consortium (DPAC,
\url{https://www.cosmos.esa.int/web/gaia/dpac/consortium}). Funding for the DPAC
has been provided by national institutions, in particular the institutions
participating in the {\it Gaia} Multilateral Agreement. Michele Scalco and Luigi Rolly Bedin acknowledge support by MIUR under PRIN program \#2017Z2HSMF.\looseness=-4

\section*{Data Availability}
All data analysed in this paper are publicly available from the \textit{Gaia} archive (\url{http://gea.esac.esa.int/archive/}). The data presented in this article may be shared on reasonable request to the corresponding author.\looseness=-4

\bibliographystyle{mnras}
\bibliography{main}

\appendix
\section{Gaia DR3 Data Selections}

\begin{centering}
\begin{figure}
\centering
 \includegraphics[width=\columnwidth]{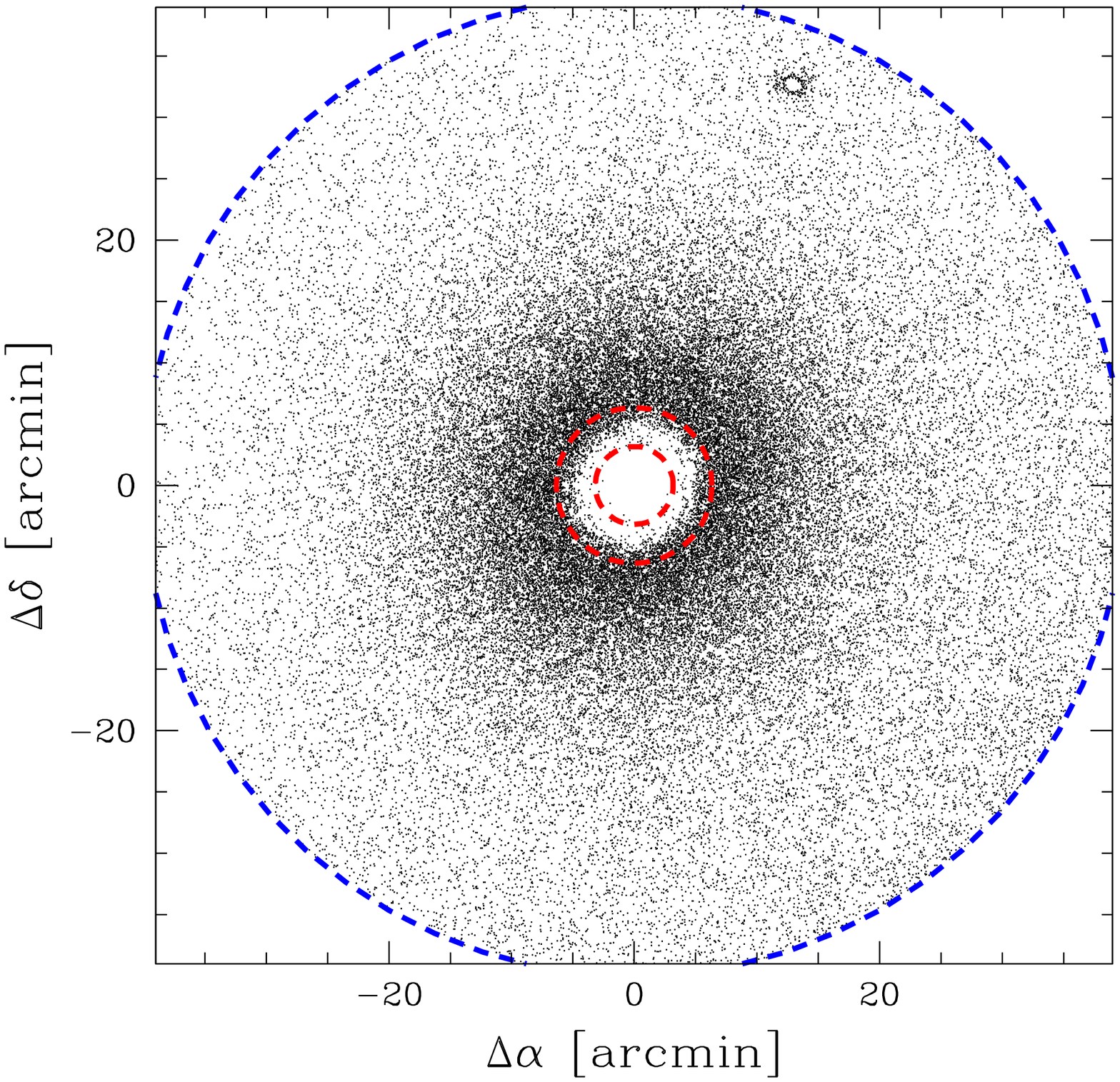}
 \caption{Spatial distribution of the selected sources for 47 Tuc. The red dashed circles represent the cluster’s half-light radius ($r_{\rm h}$ = 3.17 arcmin \citealt{1996AJ....112.1487H,2010arXiv1012.3224H}) and 2$r_{\rm h}$, while the blue dashed circle represents the extraction radius ($r_{\rm ext}$ = 40 arcmin).} 
 \label{Figure0} 
\end{figure} 
\end{centering}

For each of the three clusters we retained all sources from the \textit{Gaia} DR3 catalogue with:
\\
\\
\schema[open]{}{
\texttt{13 < G < 20},\\
\hspace*{2.5ex}\texttt{RUWE < 1.15},\\
\hspace*{2.5ex}\texttt{astrometric\_excess\_noise\_sig $\leqslant$ 2},\\
\hspace*{2.5ex}\texttt{ipd\_gof\_harmonic\_amplitude < 0.1},\\
\hspace*{2.5ex}\texttt{ipd\_frac\_multi\_peak $\leqslant$ 2},\\
\hspace*{2.5ex}\texttt{visibility\_periods\_used $\geqslant$ 10},\\
\hspace*{2.5ex}\texttt{duplicated\_source = 0},\\
\hspace*{2.5ex}\texttt{astrometric\_gof\_al < 4} and\\
\hspace*{2.5ex}\texttt{\textit{C} - \textit{f}(\textit{G}$_{\rm BP}$ - \textit{G}$_{\rm RP}$) < 3$\sigma_C$}, 
}
\\
\\
where $C$ is the \texttt{phot\_bp\_rp\_excess\_factor}, $f(x) =\sum_i a_i x^i$, with the polynomial coefficients $a_i$ taken from Table 2 of \citet{2021A&A...649A...3R} and $\sigma_C$ given by Equation (18) of \citet{2021A&A...649A...3R}.
\\
These selections are particularly severe for the very central regions of the clusters due to the limited capabilities of \textit{Gaia} in crowded and dense environments. As an example of the result of the selection procedure, we show in Fig. \ref{Figure0} the distribution of selected sources in the plane of the sky for 47 Tuc. Due to the selections, the inner part of the cluster is strongly underpopulated, with essentially no stars inside the cluster's half-light radius, $r_{\rm h}$. A similar central depletion is present in the data selected for the analysis of $\omega$ Cen and M 5.

\end{document}